# Altermagnetic Polar Metallic phase in Ultra-Thin Epitaxially Strained RuO$_2$ Films

Seung Gyo Jeong[1,†,*], In Hyeok Choi[2,†,*], Sreejith Nair[1], Luca Buiarelli[1], Bita Pourbahari[3,4], Jin Young Oh[5], Nabil Bassim[3,4], Daigorou Hirai[6], Ambrose Seo[7], Woo Seok Choi[5], Rafael M. Fernandes[8,†], Turan Birol[1], Liuyan Zhao[9], Jong Seok Lee[2,*], and Bharat Jalan[1,*]

[1]*Department of Chemical Engineering and Materials Science, University of Minnesota−Twin Cities, Minneapolis, Minnesota 55455, United States*

[2]*Department of Physics and Photon Science, Gwangju Institute of Science and Technology (GIST), Gwangju 61005, Republic of Korea*

[3]*Canadian Centre for Electron Microscopy and Department of Materials Science and Engineering, McMaster University, Hamilton, ON L8S 4L8, Canada*

[4]*Department of Materials Science and Engineering, McMaster University, Hamilton L8S 1L9, Canada*

[5]*Department of Physics, Sungkyunkwan University, Suwon 16419, Republic of Korea*

[6]*Department of Applied Physics, Graduate School of Engineering, Nagoya University, Chikusa-ku, Nagoya 464-8603, Japan*

[7]*Department of Physics and Astronomy, University of Kentucky, Lexington, Kentucky 40506, United States*

[8]*School of Physics and Astronomy, University of Minnesota−Twin Cities, Minneapolis, Minnesota 55455, United States*

[9]*Department of Physics, University of Michigan, 450 Church Street, Ann Arbor, Michigan 48109, United States*

[†] Equally contributed

[†]Current Address: *Department of Physics, University of Illinois Urbana-Champaign, Urbana, Illinois, 61801, United States*

[*]Corresponding authors: jeong397@umn.edu, jys2316@gm.gist.ac.kr, jsl@gist.ac.kr, bjalan@umn.edu

**Abstract**

Altermagnetism refers to a wide class of magnetic orders featuring magnetic sublattices with opposite spins related by rotational symmetries, resulting in non-trivial spin splitting and magnetic multipoles. However, the direct observation of the altermagnetic order parameter remains elusive. Here, by combining theoretical analysis, electrical transport, X-ray and optical spectroscopies, we establish a phase diagram in hybrid molecular beam epitaxy-grown $RuO_2/TiO_2$ (110) films, mapping symmetries along with altermagnetic/electronic/structural phase transitions as functions of film thickness and temperature. This features a novel altermagnetic metallic polar phase in epitaxially-strained 2 nm films, extending the concept of multiferroicity to altermagnets. Such a clear signature of a magnetic phase transition at ~500 K is observed exclusively in ultrathin strained films, unlike in bulk $RuO_2$ single crystals. These results demonstrate the potential of epitaxial heterostructure design to induce altermagnetism, paving the way for emergent novel phases with multifunctional properties.

**Introduction**

Magnetic order, the self-organization of electron magnetic moments, breaks time-reversal symmetry often in conjunction with translational, inversion, and rotational symmetries[1-3]. In non-compensated magnets like ferromagnets, Zeeman coupling induces an essentially uniform spin-splitting across the Brillouin zone. In contrast, conventional collinear antiferromagnets preserve spin-degeneracy in their band structures due to the combined time-reversal (TR) and translation or inversion symmetries. Recently, altermagnets have been identified as a new class of collinear compensated magnets where TR combined with rotation preserves the magnetic order[2,4]. They exhibit momentum-dependent spin-splitting with *d*-wave, *g*-wave, or *i*-wave symmetry, without a net magnetic moment[1-15]. Recent experimental studies aimed at validating this novel phase have often relied on electrical transport and angle-resolved photoemission spectroscopy (ARPES) at low temperatures[16-23]. However, direct probing of altermagnetic order parameters and coupled symmetries near the transition temperature remains challenging and thus missing.

Rutile $RuO_2$ has been one of the initial material candidates theoretically predicted to exhibit non-relativistic spin-splitting[2,16,24]. Experimentally, whether this compound displays any type of magnetism has become the center of ongoing debates. Bulk $RuO_2$ has the tetragonal 4/*mmm* ($P4_2/mnm$) structure and is expected to exhibit 4'/*mmm*' ($P4'_2/mnm'$) magnetic point (space) group hosting the antiferromagnetic order[1], where the magnetic moments point along the [001] axis. Earlier X-ray and neutron diffraction studies identified antiferromagnetic order with a small magnetic moment in bulk $RuO_2$[25-27]. However, recent muon spin rotation (µSR) spectroscopy and neutron diffraction experiments[28,29], as well as optical conductivity and some ARPES studies[30,31] have suggested a non-magnetic phase of $RuO_2$ for bulk and/or thick films. This is consistent with density-functional-theory calculations with an added onsite repulsion (DFT+$U$), which reported that stoichiometric bulk $RuO_2$ does not have a magnetically ordered ground state unless the +$U$ added is unphysically large[24], although hole-doping or vacancies could lead to altermagnetism. Meanwhile, numerous experimental studies, particularly in $RuO_2$ films, have reported properties that are consistent with the presence of altermagnetism, possibly coexisting with some weak ferromagnetism, including spin-split bands[16,32], magneto-optical (MO) effects[33], anomalous Hall effect[17,32,34], thermal spin transport[35,36], and spin Hall effects[37-42]. While it remains elusive to conclude whether $RuO_2$ is magnetic in general, experimental observations suggest that $RuO_2$ films more commonly display magnetic behavior than bulk samples do. To shed light on this issue, it is essential to establish the role of epitaxial strain in influencing altermagnetism. This highlights the critical need for systematic investigations into magnetism and its associated order parameter in high quality $RuO_2$ films as a function of strain. By modulating structural and electronic properties[43-45], strain can emerge as a key tuning parameter for the magnetic state of $RuO_2$.

To address this issue, we investigated strained $RuO_2$ films down to the nanometer thickness limit, using epitaxial $RuO_2$/$TiO_2$ (110) heterostructures grown by hybrid molecular beam epitaxy (hMBE). Crystal symmetry and magnetic order were characterized using symmetry-sensitive optical second harmonic generation (SHG) and MO effects (Fig. 1). By tracking the dependence of magnetic-SHG signal[46-49] and the MO signal[50,51], we directly observe a change from a non-magnetic point group to a magnetic point group below a transition temperature $T_c \sim 500$ K for fully-strained films with thicknesses below 4 nm, consistent with an altermagnetic state. In contrast, no such signal was observed in bulk $RuO_2$ single crystals, which is in agreement with the previously reported nonmagnetic state in bulk $RuO_2$ sample from the μSR study[29]. We attribute the unique magnetic behavior to the epitaxial strain present in ultrathin $RuO_2$ films, which is further supported by DFT calculations. Notably, we identified a novel altermagnetic polar metallic phase in epitaxially-strained 2 nm $RuO_2$ film. Our findings offer novel insights into altermagnetism in rutile compounds and highlight the potential of epitaxial design for controlling altermagnets.

**Results and Discussion**

We first discuss the crystal structure of atomically flat $RuO_2$/$TiO_2$ (110) films as a function of film thickness ($t$) (Supplementary Fig. 1). Fig. 2a schematically illustrates the strained $RuO_2$/$TiO_2$ (110) film, with lattice mismatches of $-4.7\%$ along [001] and $+2.3\%$ along [1$\bar{1}$0][52,53]. This anisotropic strain reduces the crystalline symmetry of $RuO_2$ from tetragonal $P4_2/mnm$ to orthorhombic $Cmmm$, which can be further reduced to polar $Amm2$ due to interface effects in thin films (Supplementary Fig. 2). Fig. 2b shows that X-ray diffraction (XRD) reciprocal space maps (RSMs) at the (310) plane of 4 and 17 nm $RuO_2$ films demonstrating a fully-strained state along [1$\bar{1}$0] in the 4 nm film and partial relaxation in 17 nm film. Additionally, RSMs at the (332) plane (Supplementary Fig. 3) confirm the fully-strain state along [001] in the 4 nm film and reveal the onset of strain relaxation along [001] for films with $t \geq 6$ nm. The high-angle annular dark-field scanning transmission electron microscopy of an 11 nm $RuO_2$ film exhibits local dislocations along the [001] direction due to these strain relaxations (Supplementary Fig. 4), consistent with XRD results.

To investigate the symmetries of these films, we performed rotational anisotropy (RA) SHG measurements at an oblique angle $\theta$ in a reflection geometry by rotating the azimuthal angle $\varphi$ relative to the in-plane crystal axis [1$\bar{1}$0] (Fig. 1). The polarizations of fundamental and the SHG light were set to be parallel (P) or normal (S) to the incidence plane with P/$S_{in}$-P/$S_{out}$ channels (equivalently PP, PS, SP, and SS). Fig. 2c shows the $t$-dependent evolution of SHG polar plots in the PP ($I_{SHG}^{PP}(\varphi)$, upper panels) and SS configurations ($I_{SHG}^{SS}(\varphi)$, lower panels) at $\theta = 45°$ and 560 K to focus on the crystalline symmetry characterization. The RA SHG patterns showed pronounced $t$-dependence with abrupt changes for $t \geq 4$ nm. Fig. 2d presents the angular-averaged SHG field strengths as a function of $t$, i.e., $\sqrt{I_{SHG}^{PP \text{ or } SS}}(t) = \sqrt{\sum_{\varphi} I_{SHG}^{PP \text{ or } SS}(\varphi, t)}$, exhibits a $t$-dependent enhancement with distinct slopes for $t \leq 2$ nm and $t \geq 4$ nm, suggesting a structural phase transition as a function of $t$. Consistently, dramatic changes in crystal field energy and Hall coefficients are also observed within the same thickness range (Supplementary Fig. 5).

Detailed analyses of the RA SHG patterns reveal the crystalline symmetries as a function of thickness. For strained $RuO_2$ films ($t \leq 2$ nm), $I_{SHG}^{SS}$ is negligible, while the other polarization configurations exhibit 2-fold symmetric SHG responses. For example, in the $t = 2$ nm film (Supplementary Fig. 6), RA SHG in the PP, SP, and PS configurations exhibits a 2-fold rotational anisotropy about the [110] axis and two mirrors normal to the [1$\bar{1}$0] and [001] directions. These symmetries, revealing the breaking of the mirror perpendicular to the [110] axis, correspond to the non-centrosymmetric point group *mm2*, with the electric dipole contributions (ED, blue lines in Fig. 2c) dominating the SHG response for $t \leq 2$ nm (Supplementary Notes 1, Supplementary Table 1, and Supplementary Figs. 6, 7, and 8). The $\theta$-dependence of $I_{SHG}$ further supports the $mm2$ symmetry for $t \leq 2$ nm (Supplementary Notes 2 and Supplementary Fig. 9). Furthermore, for $t \leq 2$ nm, $I_{SHG}$ increases with $t$, suggesting that ED SHG is a bulk response throughout the whole film thickness. Group theory shows that, starting from the tetragonal group $4/mmm$ of the bulk crystal, strain (or magnetic order) alone cannot induce the $mm2$ point group. Instead, a polar field along [110] is required, which can arise from interface effects or intrinsic polar symmetry breaking (Supplementary Fig. 2). Previous DFT calculations suggested that strain-induced zone-boundary lattice instabilities alone are insufficient[54], but an interfacial polar field can stabilize a polar phase throughout the RuO$_2$ film (Fig. 2e). Meanwhile, in strain-relaxed films ($t \geq$ 4 nm), the RA SHG patterns are well explained by the electric quadrupole (EQ) contribution to SHG with an underlying non-polar, centrosymmetric point group $4/mmm$ (black and grey lines in Fig. 2c), consistent with bulk RuO$_2$. As $t$ increases for $t \geq 4$ nm, both $I_{SHG}^{PP}$ and $I_{SHG}^{SS}$ increase, suggesting the bulk origin of $4/mmm$ EQ SHG.

The temperature dependence of the RA SHG shows clear evidence for time-reversal symmetry breaking in the strained RuO$_2$ films (i.e., films with $t \leq 4$ nm) below $T_c \sim 500$ K (Figs. 3a-3c). Fig. 3a compares $I_{SHG}^{PP}(\varphi)$ below $T_c$ (320 and 420 K) and above $T_c$ (540 K) for the 2 nm RuO$_2$ films. The anisotropy of $I_{SHG}^{PP}(\varphi)$ shows a significant temperature dependence with a large enhancement of the signal amplitude upon cooling, whereas $I_{SHG}^{SS}$ shows no meaningful temperature dependence, remaining below the noise level (Supplementary Figs. 10 and 11). Most importantly, the symmetry of $I_{SHG}^{PP}(\varphi)$ below $T_c$ is lower than its symmetry above $T_c$, signaling a phase transition. To verify whether there is a structural phase transition near $T_c$, we conducted temperature-dependent XRD and reflection high-energy electron diffraction (RHEED) measurements for the RuO$_2$ films (Supplementary Fig. 12). These measurements confirmed no structural transitions between 300 and 560 K. Thus, we attribute the temperature-dependent $I_{SHG}^{PP}(\varphi)$ to time-reversal symmetry breaking in strained RuO$_2$ films. For the strain-relaxed 17 nm film, SHG data shows no evidence of broken symmetry, remaining consistent with the tetragonal $4/mmm$ group across all temperatures (see Supplemental Fig. 13). We also measured bulk RuO$_2$ single crystal, showing identical RA SHG patterns between 320 K and 520 K (Supplementary

Fig. 14), consistent with μSR studies[29], highlighting the key role of epitaxial strain in the evolution of SHG signals in our 2 nm strained $RuO_2$ films.

To elucidate the origin of this broken symmetry, we plot in Fig. 3b the second-order susceptibility tensors $\Delta\chi_{PP}^i$ and $\Delta\chi_{PP}^c$ obtained from $I_{SHG}^{PP}$. Clearly, both $\Delta\chi_{PP}^i$ and $\Delta\chi_{PP}^c$ exhibit an anomaly around 500 K. Thus, the change in the symmetry of $I_{SHG}^{PP}(\varphi)$ can be attributed to nonlinear optical responses. As we explained above, at 540 K, $I_{SHG}^{PP}(\varphi)$ is mirror-symmetric for $\varphi = 0°$ and 90° (dotted lines along $[1\bar{1}0]$ and [001] in Fig. 3c), consistent with the structural point group *mm*2. However, at 320 K and 420 K, the 2-fold symmetric $I_{SHG}^{PP}(\varphi)$ shows dips at $\varphi = 60°$ and 240° (marked by the arrows) that are absent at their mirror-symmetric angular counterparts $\varphi = 120°$ and 300° (Figs. 3a and 3c). This reveals that the strained 2 nm $RuO_2$ films break two mirrors normal to the $[1\bar{1}0]$ and [001] directions, while retaining the 2-fold rotational axis along [110]. Since a structural transition was ruled out by our XRD and RHEED measurements, we consider whether a magnetic transition could lead to the broken mirrors observed experimentally. Starting with the paramagnetic group *mm*2.1′, we find that the magnetic subgroup with the highest symmetry has the same symmetries as $I_{SHG}^{PP}(\varphi)$ below $T_c$ is *m′m′*2 (here, the prime indicates time-reversal symmetry). The key point is that the *m′m′*2 magnetic point group allows for both time-invariant (*i*-type) and time-variant (*c*-type) ED contributions to SHG, which scale with even and odd powers of the magnetic order parameter, respectively. The RA SHG at 320 K and 420 K fits well with the coherent interference between the *i*-type magnetic ED SHG of *m′m′*2 (or equivalently, structural ED SHG of *mm*2) and *c*-type magnetic ED SHG of *m′m′*2, as shown by the blue line in Figs. 3a and 3c. The identification of this magnetic group is also supported by the absence of $I_{SHG}^{SS}$.

Thus, we conclude that the strained $RuO_2$ thin films undergo a magnetic transition towards a ground state with magnetic point group *m′m′*2. This is the main result of our paper. Such a magnetic group naturally emerges by considering a translationally-invariant compensated magnetic state with magnetic moments aligned along [110] (Fig. 3d). This is a primary altermagnetic state, since the magnetic moments in opposite sublattices are related by a non-symmorphic symmetry[1-3], even though a secondary ferromagnetic order parameter is also allowed by symmetry. Note that this orientation of the moments is different from that theoretically proposed in bulk $RuO_2$ crystals[4,17], where the moments point along [001] (Fig. 3e). In this case, the magnetic subgroup would be $m'm2'$, requiring mirror symmetry along [001] (Supplementary Notes 1). However, this contradicts our RA SHG data observations, which show no such mirror symmetry (green curve in Fig. 3c).

This $m'm'2$ altermagnetic state in the 2 nm-thick strained $RuO_2$ film is also confirmed by the MO measurements. Applying a magnetic-($H$-) field along the [110] direction, the magneto-optical Kerr effect (MOKE) is symmetry-allowed for both $mm2$ and $m'm'2$, while the magnetic linear dichroism (LD) signal is only permitted for $m'm'2$ (Supplementary Notes 3). To enhance the MO signal in the ultra-thin $RuO_2$ layer, we designed a superlattice consisting of alternating 2 nm $RuO_2$ and 1 nm $TiO_2$ layers with five repetitions, i.e., $RuO_2/TiO_2$ SL, showing a fully strained state in XRD (Supplementary Fig. 15). MOKE and magnetic LD measurements were conducted with alternating $H$-field ($H = \pm 300$ mT) along the out-of-plane direction (Fig. 3f), and the results for strained $RuO_2$ showed distinct temperature-dependent behaviors (Figs. 3g-3h and Supplementary Fig. 16), consistent with symmetry predictions. In Fig. 3g, we observed a mild peak-like feature at 530 K in $\sigma_{xy,1}$ of the $RuO_2/TiO_2$ SL, obtained from complex Kerr angle $\tilde{\theta}_K$, whereas that of $TiO_2$ showed negligible values with no significant temperature-dependence. Since $\sigma_{xy,1}$ reflects the $H$-field-induced magnetic moment, proportional to magnetic susceptibility[55], this peak at $T_c$ aligns with the expected Curie-Weiss law for an antiferromagnet, which peaks at the transition temperature. For the LD measurement, we pumped the film with a pulsed laser and monitored the pump-induced changes in LD ($\Delta$LD). Temperature-dependent $\Delta$LD in Fig. 3h exhibits a broad peak-like feature at ~ 540 K, consistent with $\sigma_{xy,1}$. Since $\Delta$LD is related to the thermo-LD coefficient ($d$LD$/dT$) and the pump-induced temperature increase ($\Delta T$) as $\Delta$LD ~ ($d$LD$/dT$) × $\Delta T$, cumulative integration of $\Delta$LD yields a signal directly proportional to the static LD. In the inset of Fig. 3h, the temperature-dependent LD signal increases starting around 540 K, indicating the development of an altermagnetic order parameter. These MO measurements thus provide direct evidence for a magnetic ground state in strained $RuO_2$ films, thereby validating our symmetry-based arguments.

To understand the altermagnetic state in strained $RuO_2$, we calculated the non-collinear spin density of $RuO_2$ for both $m'm'2$ and $m'm2'$ phases along the [110] and [001] spin components ($S_{[110]}$ and $S_{[001]}$) using DFT at $U = 0$ (Fig. 4a). Our calculations (See Supplementary Tables 3-6) show that epitaxial strain favors magnetism in a fully-strained $RuO_2$ film even without $U$, and the addition of $U$ only stabilizes the altermagnetic phase, in contrast to the nonmagnetic state of bulk $RuO_2$ at $U = 0$[24], highlighting the distinct magnetic properties enabled by strain in $RuO_2$. While it is also possible to stabilize a phase with the same symmetry and parallel moments, the primary-altermagnetic state is lower in energy by ~1 meV per formula unit. The integrated spin density of the primary spin component (i.e., [110] for $m'm'2$ and [001] for $m'm2'$, respectively) yields a small, but nonzero integrated magnetic moment ($\int S \neq 0$), indicating that strained $RuO_2$ is no longer a compensated magnet. The secondary components, which are perpendicular to the magnetic moment orientation (i.e., [001] for $m'm'2$ and [110] for $m'm2'$, respectively), display the magnetic octupolar symmetry with a zero integrated magnetic moment ($\int S = 0$), further illustrating their altermagnetic nature. Therefore, the calculations suggest the coexistence of altermagnetism and weak-ferromagnetism in strained $RuO_2$ films.

To investigate the interplay between electrical properties and magnetism as $t$ decreases, we performed temperature-dependent resistivity measurements. Supplementary Fig. 17 shows that $RuO_2$ films maintain metallic behavior down to $t = 2$ nm, whereas the 1.5 nm film transitions to an insulating state. The metallicity of the 2 nm $RuO_2$ film is further supported by the Drude response observed in optical spectroscopic ellipsometry (Supplementary Fig. 18). These results are summarized in a temperature-thickness phase diagram (Fig. 4b), highlighting the structural, electronic, and magnetic phases of $RuO_2$ films. Notably, a novel altermagnetic-$mm2$ (polar) metallic phase emerges at 2 nm, underscoring the versatility of $RuO_2$ epitaxial heterostructures. The coexistence of various electronic and structural phases with magnetic transitions (above room temperature) down to 1 nm demonstrates the unique advantages of $RuO_2$ thin film for potential applications (Supplementary Table 7 for a summary of previous SHG studies). The absence of transition in 0.4 nm $RuO_2$ (Supplementary Figs. 10 and 11) likely reflects the dimensional limits of this magnetism.

In summary, our multiple experimental observations and theoretical analysis uncover the versatile electronic/magnetic/structural phases in epitaxial $RuO_2$ films. We reveal the fingerprints of the altermagnetic order parameter in strained $RuO_2$ film through nonlinear- and magneto-optical responses, which is not observed in RuO$_2$ bulk single crystal. Interestingly, a structural phase transition from the centrosymmetric 4/*mmm* to the inversion-symmetry-broken *mm*2 group is realized via heterostructure design, promoting a novel altermagnetic polar phase in strained RuO$_2$ film. The observed altermagnetic order parameters open new avenues for investigating unique phenomena and applications in the realm of oxide heterostructures and altermagnetic materials.

**Methods**

**Solid-source metal-organic hMBE**

High-quality epitaxial RuO$_2$ films were synthesized using an oxide hMBE system (Scienta Omicron) on a TiO$_2$ (110) single crystalline substrate (Crystec). Solid-source metal-organic MBE has demonstrated high-quality RuO$_2$ thin films with an atomically flat surface, enabling precise and systematic analysis down to a nano-scale limit (Supplementary Fig. 1), which was never reported in any other film growth methods[56,57]. We utilized metal-organic precursor Ru(acac)$_3$, which was thermally evaporated using a low-temperature effusion cell (MBE Komponenten) set at an effusion cell temperature between 170-180 °C[52,58]. Before film growth, we performed a substrate treatment process with acetone, methanol, and isopropanol, followed by 2-hour baking at 200 °C in a load lock chamber. We further performed a 20-minute annealing in oxygen plasma at 300 °C of growth temperature before the film growth. We used radio frequency oxygen plasma operating at 250 W and an oxygen gas pressure of $5 \times 10^{-6}$ Torr. After the growth, the sample was cooled to 120 °C in the presence of oxygen plasma to prevent the potential formation of oxygen vacancies.

**Structure characterization**

High-resolution XRD measurements were performed using a Rigaku SmartLab XE and a PANalytical X'Pert XRD. We determined the crystallinity, film thickness, roughness, and strain state using reciprocal space mapping (RSM) and X-ray reflectivity and θ-2θ scans. For temperature-dependent XRD measurement, we used the sample heating stage of SmartLab XE with a spherical X-ray window (DHS-1100) in atmospheric conditions. We monitored film surfaces before, during, and after growth using in-situ reflection high-energy electron diffraction from Staib Instruments. Surface morphologies of RuO$_2$ films were further confirmed by using atomic force microscopy (Bruker Nanoscope V Multimode 8) with peakforce tapping mode.

**Optical second harmonic generation**

SHG is a suitable observational tool for investigating both structural and magnetic symmetry and their correlation with magnetic orders. SHG has the unique advantage of probing magnetic transition and magnetic order parameters by measuring temperature anomaly near $T_c$, summarized in Supplementary Table 7. We employed the SHG experiment in oblique incidence geometry with rotating the sample azimuth ($\varphi$) and incidence angle ($\theta$). The fundamental beam (800 nm) is set to be P- or S-polarized, and we monitor the P- or S-polarized second harmonic beam. The notation PP and SS represent P$_{in}$-P$_{out}$ and S$_{in}$-S$_{out}$, respectively. The incident beam is focused with a beam size of 30 μm and its power is about 50 mW. The SHG signal is measured using a photomultiplier detector (*Hamamatsu*), and we use both a short-pass and band-pass filter to eliminate the fundamental signal. We conducted all temperature-dependent SHG experiments in atmosphere conditions with the cooling process. To obtain the azimuth-dependent SHG polar pattern, we rotated the sample from 0° to 360°. For measuring the temperature-dependent SHG anisotropy pattern, we rotated the sample from 0° to 180°, and duplicated these results for the data from 180° to 360°. We have confirmed that duplicated data matches well the full-rotated data (Supplementary Fig. 19). The temperature dependence of $\Delta\chi_{PP}^{i}$ and $\Delta\chi_{PP}^{c}$ are obtained by subtracting their respective values at 560 K. The $\Delta\chi_{PP}^{i}$ and $\Delta\chi_{PP}^{c}$ corresponds to $\Delta(\chi_{zzz} + \chi_{zxx} + \chi_{zyy})$ and $\Delta(\chi_{xyz} + \chi_{xzy} + \chi_{yxz} + \chi_{yzx} - \chi_{zxy} - \chi_{zyx})$, respectively.

**Magneto-optical measurements**

We conducted field-dependent magneto-optical Kerr effect (MOKE) and linear dichroism (LD) measurements on [2 nm RuO$_2$|1 nm TiO$_2$]$_5$ superlattices and the TiO$_2$ substrate to examine magnetic properties. All magneto-optic signals were obtained under an external magnetic field of 300 mT along the out-of-plane direction ([110]), and we extracted pure magnetic signals by alternatively changing the

direction of the magnetic field. For MOKE measurements, complex Kerr angle ($\tilde{\theta}_K$) consisted of Kerr rotation angle ($\theta_K$) and Kerr ellipticity ($\eta_K$) was obtained in a polar configuration, using an 80 MHz pulsed laser (Vision-S, *Coherent*). To selectively detect the signals from $RuO_2$ layers, we used 1.6 eV laser excitation below the $TiO_2$ band gap (~3 eV). In a polar configuration, the polarization of the laser was set along the [001] direction. To measure $\tilde{\theta}_K$, we modulated the probe beam with a photo-elastic modulator (50 KHz), and the modulated signal was demodulated by a digital lock-in amplifier (*Zurich*), where the reference frequency is 50 kHz (1f) and 100 kHz (2f) for $\theta_K$ and $\eta_K$, respectively. The longitudinal conductivity ($\sigma_{xx,1}$) was obtained by ellipsometry at room temperature (Supplementary Fig. 18) and assumed to be temperature-independent. Through reflectance measurements with the 1.6 eV laser excitation, we observed a 4% decrease in the 560 K value compared to room temperature (Supplementary Fig. 20), which exhibits much smaller temperature dependence compared to the SHG and MOKE signals. Hence the room-temperature value of $\sigma_{xx,1}$ was taken for deducing the $\sigma_{xy,1}$ in Fig. 3g. For LD measurements, LD signals were obtained using a photo-elastic modulator, alternatively modulating light polarizations between two orthogonal in-plane directions. We note that pump-induced changes in LD signal ($\Delta$LD), rather than static LD signal, were measured to enhance the sensitivity to the optical anisotropy, as described in ref.[59].

**Electrical transport measurement**

Temperature-dependent longitudinal and Hall resistances were measured in the van der Pauw geometry using the DynaCool Physical Property Measurement System (PPMS, Quantum Design). Ohmic contacts were obtained using aluminum wire bonding directly on the film.

**Scanning transmission electron microscopy**

Scanning transmission electron microscopy (STEM) samples were prepared using a Focused Ion Beam Scanning Electron Microscope (FIB-SEM) equipped with a Ga ion gun (Thermo Fisher Helios 5 UC). Tungsten (W) and carbon (C) films were deposited on the area of interest before FIB milling to protect the surface from ion damage. A double-corrected HRTEM/STEM (TFS Spectra Ultra, operated at 300 keV) was used to acquire High Annular Angle Dark Feld (HAADF) images of the $RuO_2/TiO_2$ interfaces, with sensitivity towards Ru and Ti lattices. The inverse FFT technique was used to unveil structural dislocation originating from strain relaxation during atomic-scale investigations.

**Density functional theory calculation**

Non-collinear DFT calculations with spin-orbit coupling were performed within the Local Density Approximation as implemented in ABINIT and using a basis of Projector Augmented Waves with Jollet-Torrent-Holzwarth PAW type pseudo-potentials, using an energy cutoff of 550 eV[60-62]. For all calculations, the Brillouin Zone (BZ) was sampled with a grid of 16 × 16 × 20 *k*-points centered at Γ. The bulk *P4$_2$/mnm* structure was fully relaxed until the forces on each atom were less than 1 meV/Å. The strained (*Cmmm*) structure was obtained by fixing the strain on the directions [1$\bar{1}$0] by + 2.5% and [001] by – 5% with respect to the relaxed bulk structure, and then allowing the components of the lattice parameters along the [110] direction and all ionic positions in the unit cell relax. The *Cm'm'm* (*Cmm'm'*) magnetic structure was obtained by imposing the initial spin-magnetization for the two Ru atoms as antiparallel along the [001] ([110]) direction in the strained structure *Cmmm*. The first principles calculations are limited to the centrosymmetric structures where the effects of magnetic order and strain are taken into account, but the interfacial dipolar field is ignored. The spinor wave functions $\psi_{nk}$ of the electronic ground states were used to calculate the spin-magnetization in the real space unit cell,

$$S_i(r) = \sum_{nk \text{ occupied}} \psi_{nk}^\dagger(r)\, \sigma_i \psi_{nk}(r)$$

where *n* indicates the band, *k* the *k*-point in the BZ and $\sigma_i$ represent the three Pauli matrices with *i* = *x*, *y*, *z*. The three orthonormal cartesian axes (*x, y, z*) are aligned with the orthogonal primitive lattice vectors (*a, b, c*) of the original, non-strained, unit cell (in the strained case, *a* is kept aligned with *x*, while *b* acquires a small non-zero *x* component, but still lies in the *xy* plane). In Figure 4a, $S_{[001]}$ is referred to as $S_z$, while $S_{[110]} = (S_x + S_y)/\sqrt{2}$. The isosurfaces were plotted using VESTA[63]. For clarity, different iso-surface values are used for each panel, and the ratio of the isosurface values for $S_{[110]}$ to $S_{[001]}$ is ~10 for the upper panels of Fig. 4a, and it is ~$10^{-1}$ for the lower panels. As enforced by the magnetic point groups, the integral of spin magnetic moment over the entire unit cell of $S_{[001]}$ in $Cmm'm'$ (top-right) and $S_{[110]}$ in $Cm'm'm$ (bottom-left) are zero. On the other hand, the integral of spin moment over the unit cell of $S_{[110]}$ in $Cmm'm'$ (top-left) and $S_{[001]}$ in $Cm'm'm$ are allowed to be non-zero, giving rise to the ferromagnetic moments in either the [110] or [001] directions. In the absence of strain, with symmetry $P4_2/mnm$, we find that the converged ground state without added $+U$ has zero magnetic moments on the Ru ions, no matter the direction of the initial moments imposed. In the presence of strain, there is a notable difference. When we impose the moments to be anti-parallel along the [001] direction, $Cm'm'm$, the converged ground state is weak-ferromagnetic (non-compensated altermagnetic) with a net moment of $0.05\,\mu_B$ along [001] in the unit cell. On the other hand, placing initial anti-parallel moments along the direction [110], $Cmm'm'$, yields a weak-ferromagnetic with a net moment of $0.01\,\mu_B$ along [110]. Introduction of the $+U$ correction favors the formation of local moments, and depending on the direction of the spins and the value of $U$, both parallel and anti-parallel configurations of spins can be stabilized with a small energy difference, as discussed in Supplementary Tables 3-6.

**X-ray absorption spectroscopy**

X-ray absorption spectroscopy (XAS) measurements were performed at the 6A beamline of the Pohang Light Source in the total electron yield (TEY) mode for the O $K$-edge. XAS spectra were obtained at room temperature with surface normal configuration and σ polarization, where the electrical field direction of the X-rays is parallel to [1$\bar{1}$0] crystal direction of $RuO_2/TiO_2$ (110) sample. We note that TEY is the surface-sensitive probing method, having the advantage of a surface state away from the substrate.

**Spectroscopic ellipsometry**

The optical conductivity spectrum was measured using spectroscopic ellipsometry (M-2000, J. A. Woollam Co., Inc.) at room temperature with photon energies ranging from 0.73 to 6.44 eV and incident angles of 60°, 65°, and 70°. The dielectric functions of the anisotropic $RuO_2$ (110) thin films along the [1$\bar{1}$0] and [001] in-plane directions were extracted using a numerical iteration process with a biaxial anisotropic layer model (optical spectrum along [001] direction is not shown). The dielectric functions of $TiO_2$ (110) were measured using a bare substrate. The consistency between the real and imaginary parts of the dielectric functions of $RuO_2$ thin films was checked with the Kramers-Kronig relations.

**Acknowledgements**

The authors appreciate Zenji Hiroi for his assistance in the growth of $RuO_2$ single crystals conducted at the Institute for Solid State Physics, University of Toyko. Structural characterization, transport, and ellipsometry at University of Minnesota (UMN) were supported by the Air Force Office of Scientific Research (AFOSR) through Grant No. FA9550-21-1-0025 and FA9550-24-1-0169. Film synthesis (S.G.J and B.J.) was supported by the U.S. Department of Energy through DE-SC0020211, and (partly) DE-SC0024710. S.N. was supported partially by the UMN Materials Research Science and Engineering Center (MRSEC) program under Award No. DMR-2011401. Parts of this work were carried out at the Characterization Facility at UMN which receives partial support from the National Science Foundation (NSF) through the MRSEC program under Award No. DMR-2011401. L.Z. acknowledges the support from Air Force Office of Scientific Research (AFOSR) Young Investigator Program (YIP) grant no. FA9550-21-1-0065 and NSF MRSEC DMR-2309029. Electron microscopy work was performed at the Canadian Centre for Electron Microscopy a core research facility at McMaster University (also supported by NSERC (Natural Sciences and Engineering Research Council of Canada) and the Canada Foundation for Innovation). B.P. and N.B. were supported by AFOSR Grant No. FA9550-23-1-0275 for electron microscopy research. A.S. acknowledges the support of NSF Grant No. DMR-2104296. R.M.F. was supported by the AFSOR under Award No. FA9550-21-1-0423. This work was supported by the National Research Foundation of Korea (NRF) grant funded by the Korea government (MSIT) (No. 2022R1A2C2007847 (I.H.C, J.S.L), 2022M3H4A1A04074153, 2021R1A2C2011340 (W.S.C.), and RS-2023-00220471 (W.S.C.)).

**Author contributions:** S.G.J., I.H.C., J.S.L., and B.J. conceived the idea and designed the experiments. S.G.J. and S.N. grew the films by hybrid MBE under the supervision of B.J. S.G.J. characterized the

**Figure Captions**

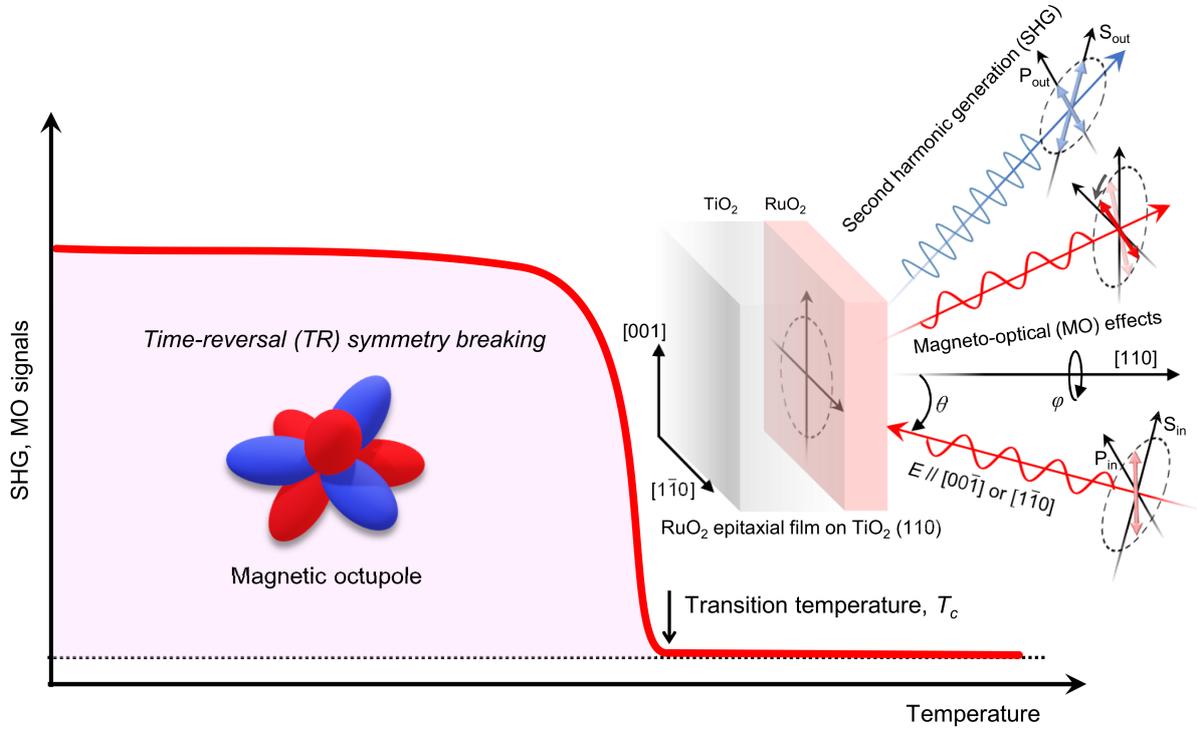

**Fig. 1. Nonlinear- and magneto-optical signal revealing the altermagnetic order parameters in RuO₂ epitaxial thin films.** Schematic illustration of the SHG and MO signals leading to the observation of an altermagnetic order parameter that breaks TR symmetry below the transition temperature $T_c$. The inset shows a schematic description of the SHG and MO geometries. The SHG signal is proportional to the square of the nonlinear susceptibility χ. The altermagnetic order parameter, corresponding to the *d*-wave spin order, is illustrated by a magnetic octupole (see also DFT results in Fig. 4).

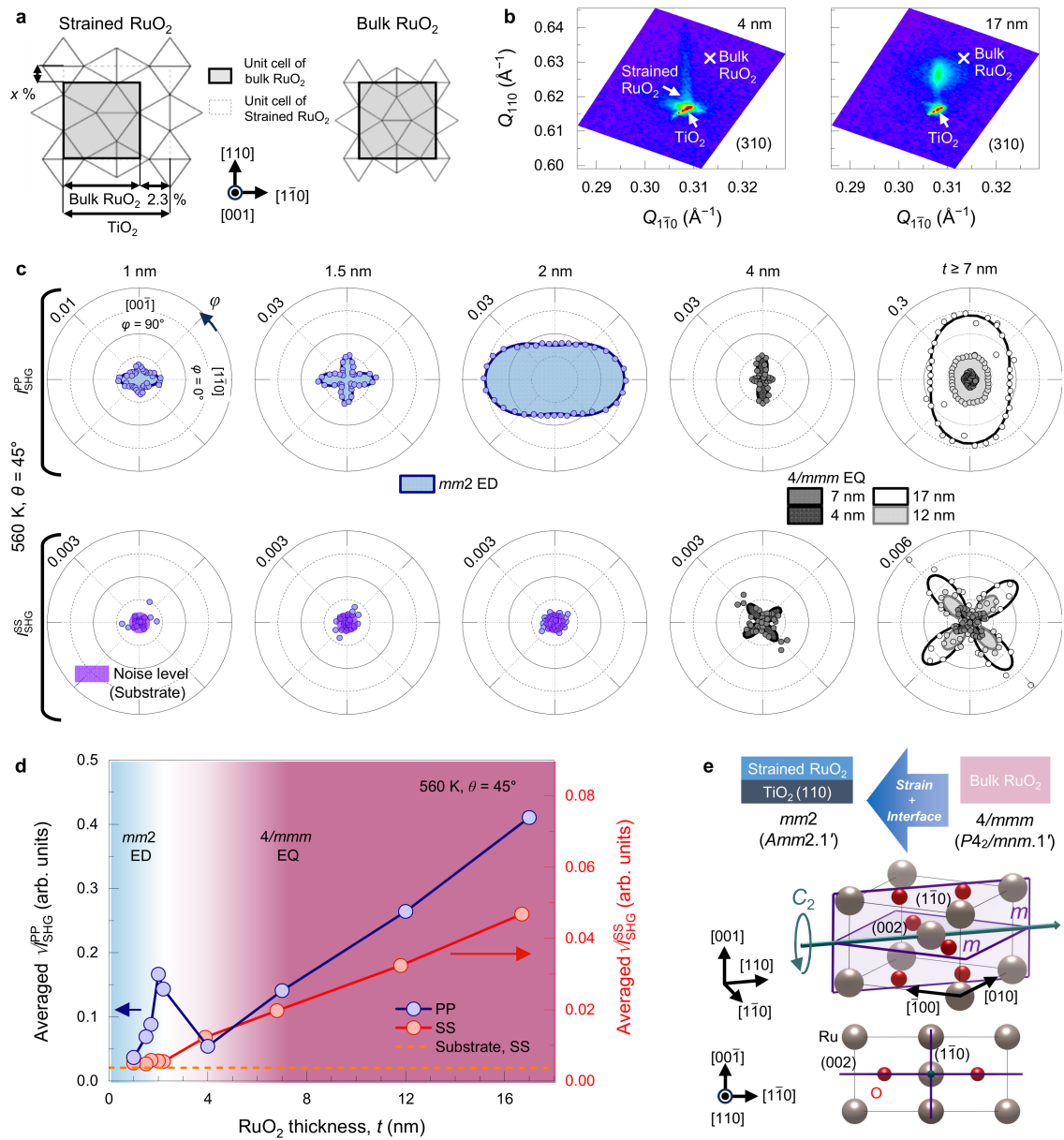

**Fig. 2. Epitaxial strain-induced crystal symmetry evolution of RuO$_2$/TiO$_2$ (110) films. a**, Schematics of strain relations for RuO$_2$/TiO$_2$ (110) using bulk lattice values. $x$ % is the lattice elongation along the [110] direction owing to the epitaxial strain. The grey square indicates the tetragonal unit cell of bulk 4/$mmm$ RuO$_2$. **b**, RSM XRD near (310) Bragg diffraction of TiO$_2$ for 4 and 17 nm RuO$_2$ films. **c**, $I_{SHG}(\varphi)$ patterns with PP (upper panels) and SS polarization (lower panels) for RuO$_2$/TiO$_2$ (110) with different thickness $t$. The colored background and lines represent model fittings (details are in the main text). The noise level (purple) in $I_{SHG}^{SS}$ estimated from the TiO$_2$ substrates. **d**, $t$-dependent averaged $\sqrt{I_{SHG}}$ values for PP and SS polarizations at 560 K. The blue and red backgrounds represent regions where strained $mm$2 films (below 2 nm) and strain-relaxed 4/$mmm$ (above 4 nm) are observed, respectively. **e**, Conceptual illustration of structural phase transition from bulk RuO$_2$ to strained RuO$_2$/TiO$_2$ (110) thin films due to the combination of strain and interface effects. The bottom panel schematically shows two mirror planes ($m$) and a 2-fold rotation axis ($C_2$) for the rutile $mm$2 structure.

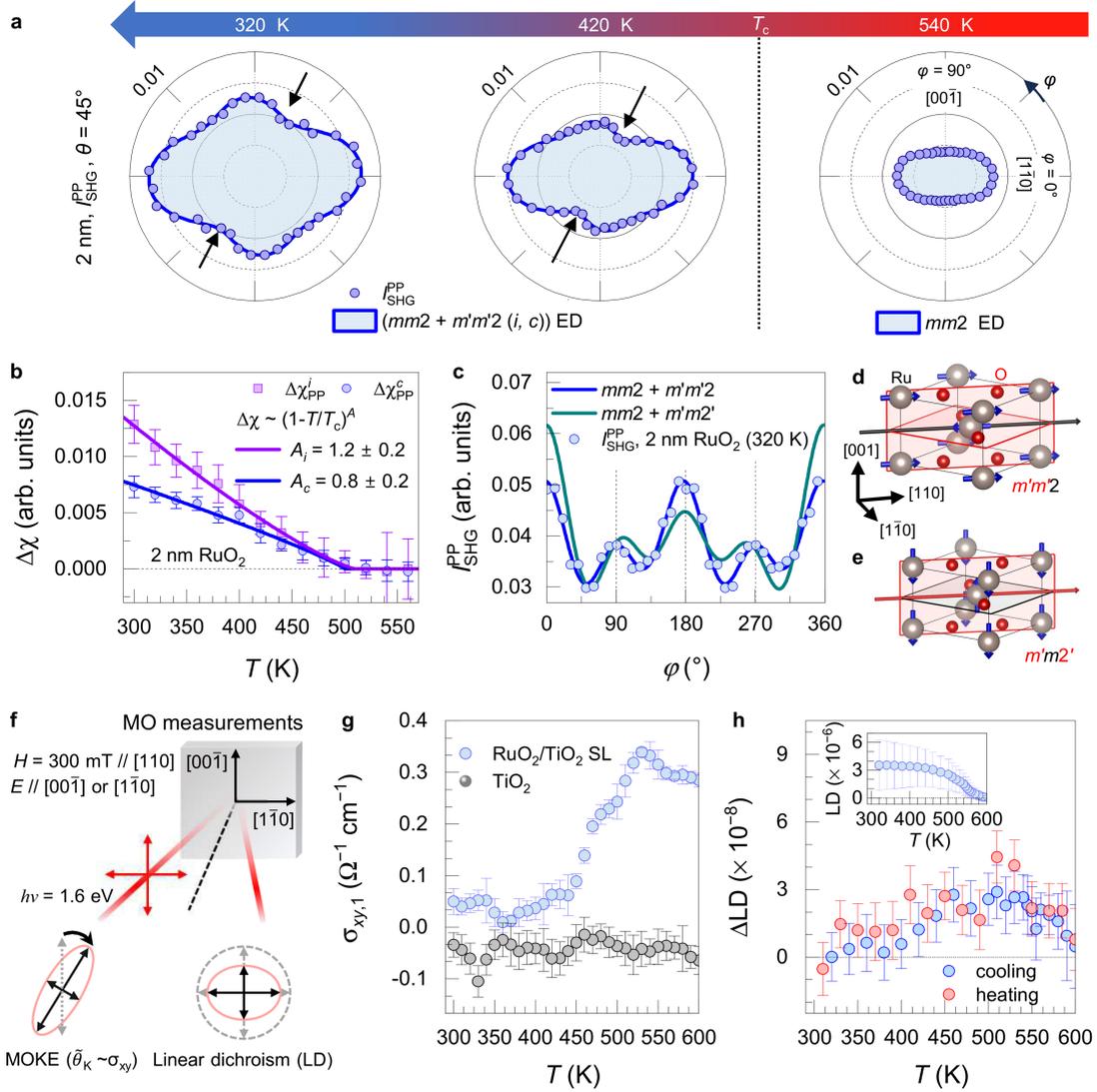

**Fig. 3. Altermagnetic phase transition in strained RuO₂ heterostructures. a**, $I_{SHG}(\varphi)$ pattern evolution for the 2 nm film with PP polarization for 320, 420, and 540 K. The arrows denote the appearance of two dips at $\varphi = 60°$ and $240°$ below $T_c$. **b**, Temperature-dependent $\Delta\chi^i_{PP}$ and $\Delta\chi^c_{PP}$ for 2 nm. The solid lines are guides to the eye obtained from the phenomenological functional forms, $(1 - T/T_c)^A$, where $T$ and $A$ are the temperature and exponent parameters, respectively. $A_i$ and $A_c$ are exponent parameters for $\Delta\chi^i_{PP}$ and $\Delta\chi^c_{PP}$, respectively. **c**, $I_{SHG}(\varphi)$ pattern for the 2 nm film at 320 K assuming two different symmetry groups. The vertical dashed line indicates the directions perpendicular to the mirrors. **d** and **e**, Schematics of the mirror planes and 2-fold rotation axis in the magnetic point group resulting from magnetic moments oriented along **d**, [110] ($m'm'2$) and **e**, [001] ($m'm2'$, lower) cases. The antichronous rotation and/or mirror elements are marked with a prime and red color. **f**, MO experiments with 300 mT of out-of-plane $H$-field and 1.6 eV of excitation energy laser. **g** and **h**, Temperature-dependent (**g**) $\sigma_{xy,1}$ and (**h**) $\Delta$LD for the 2 nm RuO₂ SL. Here, $x$ and $y$ are [001] and [1$\bar{1}$0] directions, respectively. The inset of (**h**) shows a temperature-dependent static LD signal obtained from the $\Delta$LD results during cooling. Error bars of $\sigma_{xy,1}$ and $\Delta$LD were calculated from the standard deviation of 50 data points.

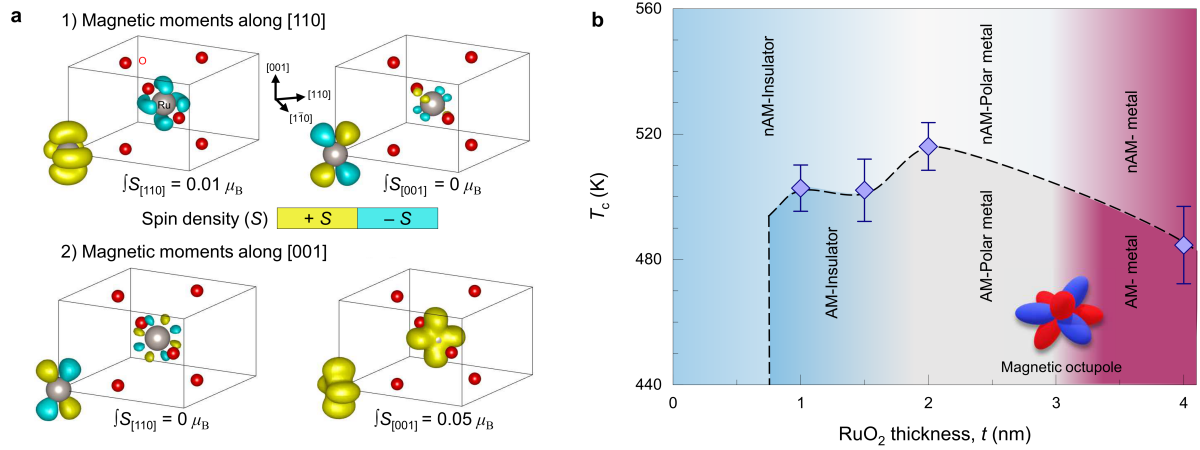

**Fig. 4. Altermagnetic phase diagram of ultra-thin RuO₂ films. a**, Calculated spin density of the [110] ($S_{[110]}$, left) and [001] spin components ($S_{[001]}$, right) for the cases of: 1) magnetic moments along [110] ($m'm'2$ magnetic point group) and 2) magnetic moments along [001] ($m'm2'$ magnetic point group). **b**, Temperature-thickness phase diagram of ultra-thin RuO₂ films. The blue symbols were obtained from the PP configurations of the SHG measurements. The inset illustrates schematically the magnetic octupole in altermagnetic RuO₂.